\documentclass[conference]{IEEEtran}
\IEEEoverridecommandlockouts
\usepackage{cite}
\usepackage{amsmath,amssymb,amsfonts}
\usepackage{algorithmic}
\usepackage{graphicx}
\usepackage{textcomp}
\usepackage[dvipsnames]{xcolor}
\usepackage{booktabs}
\def\BibTeX{{\rm B\kern-.05em{\sc i\kern-.025em b}\kern-.08em
    T\kern-.1667em\lower.7ex\hbox{E}\kern-.125emX}}

\usepackage{multirow}
\usepackage{comment}
\usepackage{subcaption}
\usepackage{tikz}
\usetikzlibrary{quantikz2}
\usepackage{amsmath}
\usepackage{url}
\usepackage{algorithm}

\begin{document}

\title{Time-Aware Qubit Assignment and Circuit Optimization for Distributed Quantum Computing

\thanks{© 2025 IEEE.  Personal use of this material is permitted.  Permission from IEEE must be obtained for all other uses, in any current or future media, including reprinting/republishing this material for advertising or promotional purposes, creating new collective works, for resale or redistribution to servers or lists, or reuse of any copyrighted component of this work in other works.

This work is sponsored in part by the Bavarian Ministry of Economic Affairs, Regional Development and Energy as part of the 6GQT project and J.S. acknowledges support from the German Federal Ministry for Economic Affairs and Climate Action through the funding program ``Quantum Computing – Applications for the industry'' based on the allowance ``Development of digital technologies'' (contract number: 01MQ22008A).}}

\author{\IEEEauthorblockN{Leo Sünkel}
\IEEEauthorblockA{\textit{Institute for Informatics} \\
\textit{LMU Munich}\\
Munich, Germany \\
leo.suenkel@ifi.lmu.de}
\and
\IEEEauthorblockN{Jonas Stein}
\IEEEauthorblockA{\textit{Institute for Informatics} \\
\textit{LMU Munich}\\
Munich, Germany}
\and
\IEEEauthorblockN{Maximilian Zorn}
\IEEEauthorblockA{\textit{Institute for Informatics} \\
\textit{LMU Munich}\\
Munich, Germany}
\and
\IEEEauthorblockN{Thomas Gabor}
\IEEEauthorblockA{\textit{Institute for Informatics} \\
\textit{LMU Munich}\\
Munich, Germany} 
\and
\IEEEauthorblockN{Claudia Linnhoff-Popien}
\IEEEauthorblockA{\textit{Institute for Informatics} \\
\textit{LMU Munich}\\
Munich, Germany}
}

\maketitle

\begin{abstract}
The emerging paradigm of distributed quantum computing promises a potential solution to scaling quantum computing to currently unfeasible dimensions. While this approach itself is still in its infancy, and many obstacles must still be overcome before its physical implementation, challenges from the software and algorithmic side must also be identified and addressed. For instance, this paradigm shift requires a new form of compiler that considers the network constraints in general as well as phenomena arising due to the nature of quantum communication. In distributed quantum computing, large circuits are divided into smaller subcircuits such that they can be executed individually and simultaneously on multiple QPUs that are connected through quantum channels. As quantum communication, for example, in the form of teleportation, is expensive, it must be used sparingly. We address the problem of assigning qubits to QPUs to minimize communication costs in two different ways. First by applying time-aware algorithms that take into account the changing connectivity of a given circuit as well as the underlying network topology. We define the optimization problem, use simulated annealing and an evolutionary algorithm and compare the results to graph partitioning and sequential qubit assignment baselines. In another approach, we propose an evolutionary-based quantum circuit optimization algorithm that adjusts the circuit itself rather than the schedule to reduce the overall communication cost. We evaluate the techniques against random circuits and different network topologies. Both evolutionary algorithms outperform the baseline in terms of communication cost reduction. We give an outlook on how the approaches can be integrated into a compilation framework for distributed quantum computing.

\end{abstract}


\section{Introduction}
The field of quantum computing has been rapidly advancing in recent years; however, the eagerly sought after quantum advantage, or even practical applicability for that matter, are still in the distant future as many obstacles are yet to be overcome. One of these obstacles is the scalability problem. While current and near-term quantum computers contain hundreds or thousands of qubits, many more are required to achieve an advantage in meaningful problems, and if error correction is to be implemented, many more are required still. One approach that has been suggested to solve the scalability problem is to connect multiple quantum computers through quantum channels that allow the use of quantum communication to leverage the power of many smaller devices to solve large problems \cite{caleffi2018quantum,awschalom2021development,cuomo2020towards}. 
However, this approach does not come without its own set of problems and challenges, and the required technology is also in its infancy. One of the problems is that quantum communication, e.g., via teleportation, requires entangled qubits in the form of bell pairs \cite{bennett1993teleporting}, and entanglement may need to be established over large distances, which can, for instance, be achieved by placing quantum repeaters throughout the network \cite{briegel1998quantum}. As establishing and maintaining entanglement is expensive, the consumption of bell pairs should be minimized as much as possible. This leads to the topic of qubit assignment in distributed quantum computing, where the aim is to assign qubits to available QPUs in the network such that overall communication cost is minimized, that is, the task is formulated as an optimization problem in which the qubit assignment can be optimized by employing various classical optimization algorithms, which is the main topic of this paper.

Therefore, we evaluate techniques in which the temporal dimension of the circuit's connectivity, i.e., multi-qubit gates, as well as the network specific topology is taken into account in the optimization. More specifically, the approaches are able to optimize when teleportation and when remote-CX gates should be applied as the circuit progresses over time. The time- and topology-aware methods include an evolutionary algorithm and a simulated annealing approach. However, we also propose an alternative technique in which the circuit itself is optimized rather than the schedule. That is, we introduce a further evolutionary algorithm inspired by techniques from quantum circuit optimization algorithms that modifies a given circuit such that a given objective is maximized. We compare our approaches to different baselines, including graph partitioning and a sequential assignment strategy and show that both evolutionary algorithms are able to outperform the baselines by significantly reducing the communication costs. 

This paper is structured as follows: We recap the fundamentals of distributed quantum computing and derive the problem statement in Section \ref{sec:background}. Related work is discussed in Section \ref{sec:related_work}, and we introduce our approaches in depth in Section \ref{sec:our_approach}. The experimental setup is discussed in Section \ref{sec:experimental_setup} and the results are presented in Section \ref{sec:numerical_results}. A discussion is given in Section \ref{sec:discussion} and we conclude in Section \ref{sec:conclusion}.

\begin{figure*}[tb]
    \centering
    \begin{subfigure}[t]{0.45\textwidth}  
        \begin{equation*}
            \boxed{
                \begin{quantikz}
                   \lstick{\ket{\psi}} & \ctrl{1} & \gate{H} & \meter{} \wire[d][2]{c} && \\
                    \lstick[2]{\ket{\Phi_+}} & \targ{} & &&  \meter{} \wire[d][1]{c} & \\
                     & & & \gate{X} & \gate{Z} & \ket{\psi}
                \end{quantikz}
            }
        \end{equation*}
        \vspace{0.9cm}
        \caption{The teleportation protocol \cite{nielsen2010quantum}. The state is teleported, i.e., transferred from the top to the bottom qubit thereby destroying it in its original location.}
        \label{fig:teleportation_protocol}
     \end{subfigure}
     \hfill 
     \begin{subfigure}[t]{0.45\textwidth}  
        \centering
        \begin{equation*}
            \boxed{
                \begin{quantikz}
                    \lstick{\ket{\psi_0}} & \ctrl{1} & & & \gate{Z} & \\
                    \lstick[2]{\ket{\Phi_+}} & \targ{} & & \meter{} \wire[d][2]{c} & & \\
                     & \ctrl{1} & \gate{H} && \meter{} \wire[u][2]{c} & \\
                    \lstick{\ket{\psi_1}} & \targ{} & & \gate{X} & &
                \end{quantikz}
            }
        \end{equation*}
        \caption{Remote CX gate where $\psi_0$ is the control and $\psi_1$ the target qubit \cite{caleffi2022distributed}.}
        \label{fig:remote_cx_protocol}
     \end{subfigure}
     \caption{The teleportation and remote CX protcols.}
\end{figure*}
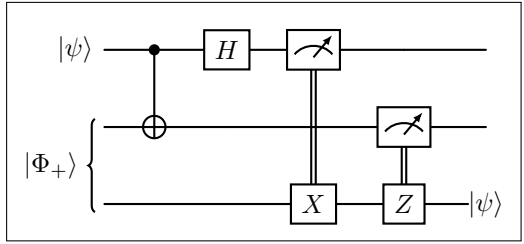
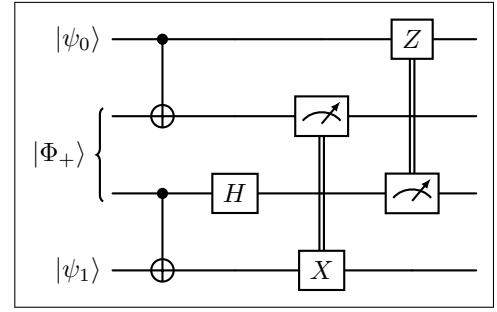

\section{Background and Problem Statement}\label{sec:background}
\subsection{Distributed Quantum Computing}
Quantum communication networks may allow many new forms of applications, including distributed quantum computing (DQC)\cite{caleffi2022distributed}, a form of quantum computing in which multiple quantum computers are connected via quantum channels with the ability to exchange quantum states using specific protocols such as teleportation, thus enabling them to collectively perform quantum computations, i.e., execute quantum circuits. This would, for instance, allow the execution of large quantum circuits that contain too many qubits for a single quantum computer, by leveraging the power of multiple smaller ones. In a quantum network, entanglement is the key resource as it is used for communication and numerous protocols, perhaps most famously \textit{quantum teleportation}. Thus, a key element of such a network is the \textit{bell state}, a state that is maximally entangled. 
As quantum nodes (QPUs) can be located far apart, certain protocols must be performed by \textit{quantum repeaters} \cite{briegel1998quantum} to enable large-distance communication. These quantum repeaters can be placed between QPUs throughout the network. For instance, \textit{entanglement swapping} can be utilized to entangle two qubits that have not interacted with each other before \cite{briegel1998quantum}. To increase the entanglement fidelity of entangled pairs, \textit{entanglement distillation} or \textit{entanglement purification} \cite{briegel1998quantum} are protocols that are able to turn $n$ weakly entangled qubits into $m$ entangled qubits with higher fidelity, given $m < n$. However, these operations are expensive and since entanglement is a valuable resource, it should be used sparingly, which leads to the main topic of this work, i.e., the minimization of communication between QPUs. Moreover, using the DQC paradigm, a large quantum circuit can be subdivided into several subcircuits, each to be executed by participating QPUs in the network. When a multi-qubit gate (e.g., CX) is used, where the target qubit is in one QPU while the control qubit is located in another, the QPUs must communicate with each other before the gate can be executed. In this situation, a qubit can be teleported such that the qubits reside in the same QPU or a remote CX gate can be executed. The paradigm shift from monolithic to distributed computation requires not only the hardware foundations for quantum communication, on the software and algorithmic side new problems and challenges also emerge. This includes new forms of compilers \cite{ferrari2021compiler,cuomo2023optimized}, and the algorithms discussed in this work can be seen as part of a larger software stack in a compilation framework for DQC. More specifically, the methods in this work compute a schedule that determines the location of each qubit at every time step of the respective circuit while minimizing the number of required remote operations. Based on this schedule, the subcircuits can be extracted and distributed to the respective QPUs in the network.

\subsection{Problem Statement}
\begin{figure}[tb]
    \centering
    \begin{equation*}
        \begin{quantikz}
            \lstick{\ket{\psi_0}} & \ctrl{1}\gategroup[4, steps=1, style={rounded corners, fill=TealBlue!20}, background]{TS 0} & & \gate{H}\gategroup[4, steps=1, style={rounded corners, fill=TealBlue!20}, background]{TS 1} && \gate{X}\gategroup[4, steps=1, style={rounded corners, fill=TealBlue!20}, background]{TS 2} & & \ctrl{3}\gategroup[4, steps=2, style={rounded corners, fill=TealBlue!20}, background]{TS 3} & \qw & \\
            \lstick{\ket{\psi_1}} & \targ{} & & \ctrl{1} && \gate{H} & & & \gate{H} & \qw \\
            \lstick{\ket{\psi_2}} & \gate{H} & & \targ{} && \ctrl{1} & & & \gate{H} & \qw \\
            \lstick{\ket{\psi_3}} & \gate{X} & &  \gate{H} && \targ{} & & \targ{} & & \qw 
        \end{quantikz}
    \end{equation*}
    \caption{Example of a random quantum circuit where time steps (TS) are marked.}
    \label{fig:example_circuit_with_time_steps}
\end{figure}
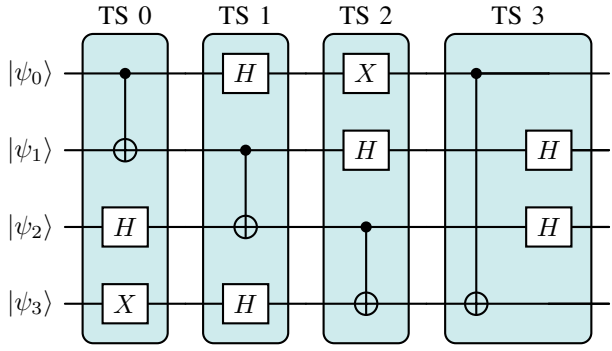

\begin{figure*}[tb]
    \centering
    \begin{subfigure}[c]{0.45\textwidth}  
    \begin{equation*}
        \begin{quantikz}
            \lstick{\ket{\psi_0}} & \gate{H} & \ctrl{1} & \gate{H} & \ctrl{1} & \rstick[2]{QPU 0} \qw \\
            \lstick{\ket{\psi_1}} & \gate{SX} & \targ{} & \ctrl{1}\gategroup[2, steps=1, style={rounded corners, fill=red!20}, background]{} & \targ{} & \qw \\
            \lstick{\ket{\psi_2}} & \ctrl{1} & \gate{X} & \targ{} & \ctrl{1} & \rstick[2]{QPU 1}\qw \\
            \lstick{\ket{\psi_3}} & \targ{} & \gate{H} & \gate{R_Z(\theta)} & \targ{} & \qw
        \end{quantikz}
    \end{equation*}
    \caption{In this example, qubits $\psi_0$ and $\psi_1$ are assigned to QPU 0 whereas qubits $\psi_2$ and $\psi_3$ to QPU 1. With this configuration, a single remote operation (marked red) is required.}
    \label{fig:remote_cx_in_qc}
    \end{subfigure}
    \begin{subfigure}[c]{0.45\textwidth} 
    \begin{equation*}
        \begin{quantikz}
            \lstick{\ket{\psi_0}} & \gate{H}\gategroup[1, steps=6, style={rounded corners, fill=red!20}, background]{} & \ctrl{1} & \gate{H} & & \ctrl{3} & \gate{R_Z(\theta)} \qw \\
            \lstick{\ket{\psi_1}} & \gate{SX}\gategroup[1, steps=2, style={rounded corners, fill=red!20}, background]{} & \targ{} & \ctrl{1}\gategroup[1, steps=4, style={rounded corners, fill=blue!20}, background]{} & \gate{H} & & \targ{} \qw \\
            \lstick{\ket{\psi_2}} & \ctrl{1}\gategroup[1, steps=6, style={rounded corners, fill=blue!20}, background]{} & \gate{X} & \targ{} & \gate{H} & & \ctrl{-1} \qw \\
            \lstick{\ket{\psi_3}} & \targ{}\gategroup[1, steps=2, style={rounded corners, fill=blue!20}, background]{} & \gate{H} & \gate{R_Z(\theta)}\gategroup[1, steps=4, style={rounded corners, fill=red!20}, background]{} & &  \targ{} & \gate{R_Z(\theta)} \qw
        \end{quantikz}
    \end{equation*}
    \caption{In this example, the qubits marked red are assigned to QPU 0 and qubits marked blue to QPU 1. In this scenario, qubits $\psi_1$ and $\psi_3$ are teleported to QPUs 1 and 0 respectively when they are needed to perform operations locally.}
    \label{fig:teleportation_in_qc}
    \end{subfigure}
    \caption{Comparison between an assignment where qubit locations remain static, i.e., in the same QPU (left) and dynamic where qubits can change their location throughout the execution over time (right).}
\end{figure*}
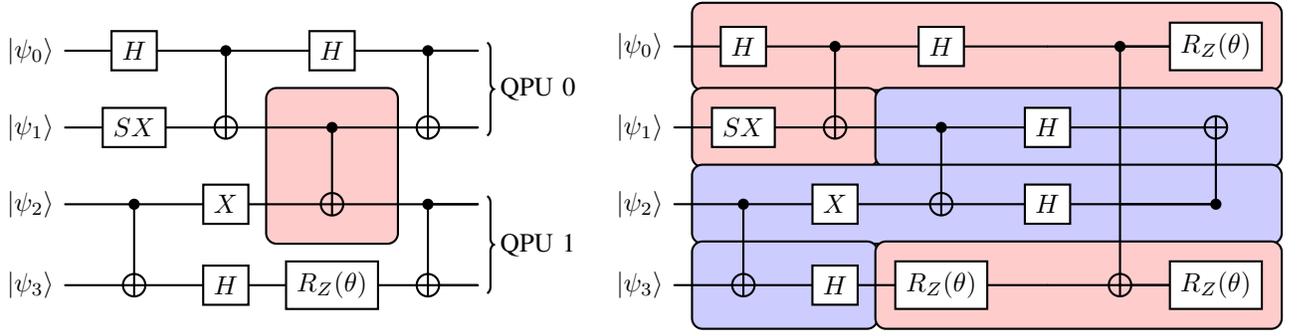

Entanglement is a valuable resource in quantum communication networks and a necessary ingredient for quantum teleportation and remote CX gates. When a large circuit is divided into smaller subcircuits, where each should be executed by a different QPU in the network, a number of multi-qubit gates (e.g. CX) may need to be executed where the involved qubits are located on different QPUs. To resolve this situation, (1) a remote CX can be executed, alternatively (2) the qubits can be teleported such that they reside in the same QPU. The teleportation protocol is shown in Fig. \ref{fig:teleportation_protocol} and the remote CX protocol in Fig. \ref{fig:remote_cx_protocol}. Both approaches consume a bell pair; thus it is crucial to minimize the number of non-local operations. Note that we solely focus on CX as multi-qubit gates in this paper. However, the methods can be adjusted to incorporate other multi-qubit gates as well.
For each circuit, a qubit schedule (i.e., assignment) can be defined that specifies the location (i.e., QPU) of each qubit for every step of the computation from which the number of required non-local operations can be extracted for each configuration \cite{sunkel2024applying}, an example circuit with corresponding schedule is shown in Fig. \ref{fig:example_circuit_with_time_steps} and Table \ref{tab:example_solution} respectively. The problem formulation is based on \cite{sunkel2024applying}.
In this two-dimensional list, each qubit is assigned a QPU for every time step, i.e., the solution representation can be viewed as a schedule that stores the location of every qubit at all times. More specifically, each row corresponds to a qubit and each column to a time step. QPUs are indexed starting from 0. A cell then contains an integer representing the QPU to which a qubit is assigned at that time. Thus, a change from one cell to the next can be viewed as a teleportation, and when two qubits involved in the same multi-qubit gate at a time step are assigned to two different QPUs a remote gate is executed. This enables the algorithm to use both teleporation as well as remote CX execution and apply each method when appropriate in order to minimize the overall communication. In some situations, it might be more efficient to execute a remote CX instead of teleporting (e.g., as shown in Fig. \ref{fig:remote_cx_in_qc}) whereas in other scenarios teleporting the qubit might be the better choice. Ideally, the optimization algorithm will balance the application of each protocol to determine the optimal schedule.

\begin{table}[b]
    \caption{Individual solution representation where each cell corresponds to $x_{i,t}$ (i.e., a QPU assignment)in the fitness formulation above.}
    \label{tab:example_solution}
    \centering
    \begin{tabular}{|l|c|c|c|c|}
    \hline
    & \multicolumn{4}{c|}{\textbf{Time steps}} \\
    \hline 
    \textbf{Qubits} & $t_0$ & $t_1$ & $t_2$ & $t_3$   \\ \hline
      $q_0$ & 0 & 0 & 0 & 0  \\ \hline
      $q_1$ & 0 & 1 & 0 & 1  \\ \hline
      $q_2$ & 1 & 1 & 1 & 1  \\ \hline
      $q_3$ & 1 & 0 & 1 & 0  \\ \hline
    \end{tabular}
\end{table}

Based on the schedule, we can formulate the following cost function consisting of three components. The first component $A$ incorporates the distance between the nodes (i.e., QPUs) of the control and target qubits of each CX gate. 

\begin{equation}
    A = \sum_{t=0}^{T-1} \sum_{\substack{(i, j) \in CX_t}}\textnormal{dist}(x_{i, t}, x_{j, t})
\end{equation}

\noindent where $T$ is the number of time steps and $CX_t$ are all CX gates at time $t$ while $x_{i,t}$ is QPU assignment (i.e, an integer) of qubit $i$ at time step $t$; $\textnormal{dist}$ is the length of the path with the shortest distance between nodes measured in the number of hops. This should encourage the assignment of qubits that often interact with each other in close proximity. Note that $x_{i,t}$ corresponds to a specific cell in the schedule shown in Tab. \ref{tab:example_solution}. The next component $B$ adds the distance when a qubit's location changes from one time step to the next, i.e, if a qubit is teleported to a different QPU:

\begin{equation}
    B = \sum_{t=0}^{T-1} \sum_{q=0}^{Q-1} \textnormal{dist}(x_{q, t}, x_{q, t+1})
\end{equation}

\noindent where $T$ is the number of time steps and $Q$ the number of qubits.
The third component $C$ enforces the capacity constraint, i.e., it adds a penalty if a QPU is assigned more qubits at a particular time step than its capacity allows:

\begin{equation}
    C = \sum_{t=0}^{T-1}  \sum_{n=0}^{N-1} g  \biggl( \biggl( \sum_{q=0}^{Q-1} [x_{q,t} = n] \biggl) - C_n \biggl)
\end{equation}

\noindent where $T$ is the number of time steps, $N$ the number of QPUs, $Q$ the number of qubits, and $C_n$ is the capacity of QPU (or node) $n$, and $g$ is defined as:

\begin{equation}
    g(x) = 
    \begin{cases}
        0 & \text{if } x \leq 0, \\
        \lambda & \text{else}.
    \end{cases}
\end{equation}

\noindent where $\lambda$ is a penalty value (set to 100 in the experiments) punishing assignments that violate the capacity constraint.

The overall cost function $f$ is then defined as:

\begin{equation}\label{eq:communication_cost}
    f(x) = A + B + C
\end{equation}

In summary, the fitness function adds the distance of the QPUs to the cost when an assigned qubit changes from one time step to the next (i.e., for each change in the row of the solution matrix above). If two qubits that are part of the same CX are assigned two different QPUs at that particular time step, a penalty (i.e., the distance between them) is also added to the cost. Finally, if a QPU is assigned more qubits than is allowed by its capacity, a penalty term $\lambda$ is added to the cost.

\section{Related Work} \label{sec:related_work}
Various approaches have been proposed to the problem of minimizing the number of teleportations, i.e., communication, circuit partitioning, as well as compiling in DQC, including EAs \cite{houshmand2020evolutionary,sunkel2024applying,burt2024generalised}, hypergraph partitioning \cite{andres2019automated}, dynamic programming \cite{davarzani2020dynamic} and a reinforcement learning based compiler \cite{promponas2024compiler}. The problem of distributing quantum circuits in a quantum network with arbitrary topology using cat-entanglement or teleportation as a means of communication is considered in \cite{sundaram2022distribution}. They propose a heuristic approach to solve a specialized version of the problem, and two algorithms for the general problem. The qubit allocation problem is also discussed in \cite{mao2023qubit} where the authors apply local search and hybrid simulated annealing algorithms on various network topologies. In \cite{dadkhah2022reordering}, an approach in which a circuit is reordered and transferred to a graph is proposed. Subsequently, tabu search combined with a genetic algorithm are applied to minimize the teleportation cost. In \cite{bandic2023mapping}, a related problem of mapping quantum circuits to QPUs in multi-core settings while minimizing communication is discussed. The authors propose a method in which the problem is formulated as a QUBO. In particular, they formulate the qubit assignment problem as a graph partitioning problem. Compilers for DQC are discussed in \cite{cuomo2023optimized,ferrari2021compiler,ferrari2023modular}.

\section{Approaches} \label{sec:our_approach}
In this section, we first discuss algorithms that optimize qubit assignment, that is, the graph partitioning approach, simulated annealing, and an evolutionary algorithm. We then propose our approach, in which the circuit itself is optimized rather than the schedule.

\subsection{Optimizing the Qubit Assignment}

\subsubsection{\textbf{Simulated Annealing}}
Simulated annealing (SA) \cite{kirkpatrick1983optimization} is a metaheuristic algorithm that can be applied to a range of optimization problems. It is an iterative approach and the main components are temperature, cooling rate, acceptance threshold, and a function to create ``neighbor'' (i.e., similar) solutions; a high-level overview of the algorithm is shown in Alg. \ref{alg:simulated_annealing}. Starting with an initial solution, a neighbor is created by adjusting the current solution. This might increase or decrease the fitness, i.e., cost of the solution. Whether a neighbor is accepted as the new solution depends on the current temperature and acceptance threshold and can be calculated as follows \cite{kirkpatrick1983optimization}:

\begin{equation}
    p(c, n, t) = \begin{cases}
        1 & \text{if } n < c,  \\
        \textnormal{exp}(-(n - c)/t) & \text{else}
    \end{cases} 
\end{equation}
\noindent where $c$ is the cost of the current solution, $n$ of the neighbor, and $t$ the current temperature.

As the algorithm progresses, the temperature gradually decreases according to a temperature or annealing schedule. With higher temperatures, the probability of accepting worse solutions is higher, allowing the algorithm to explore the solution space more freely. However, as the temperature decreases, so does the probability of accepting worse solutions, and thus the algorithm will explore less and focus on exploitation.

\begin{figure}[tb]
    \centering
    \includegraphics[scale=0.4]{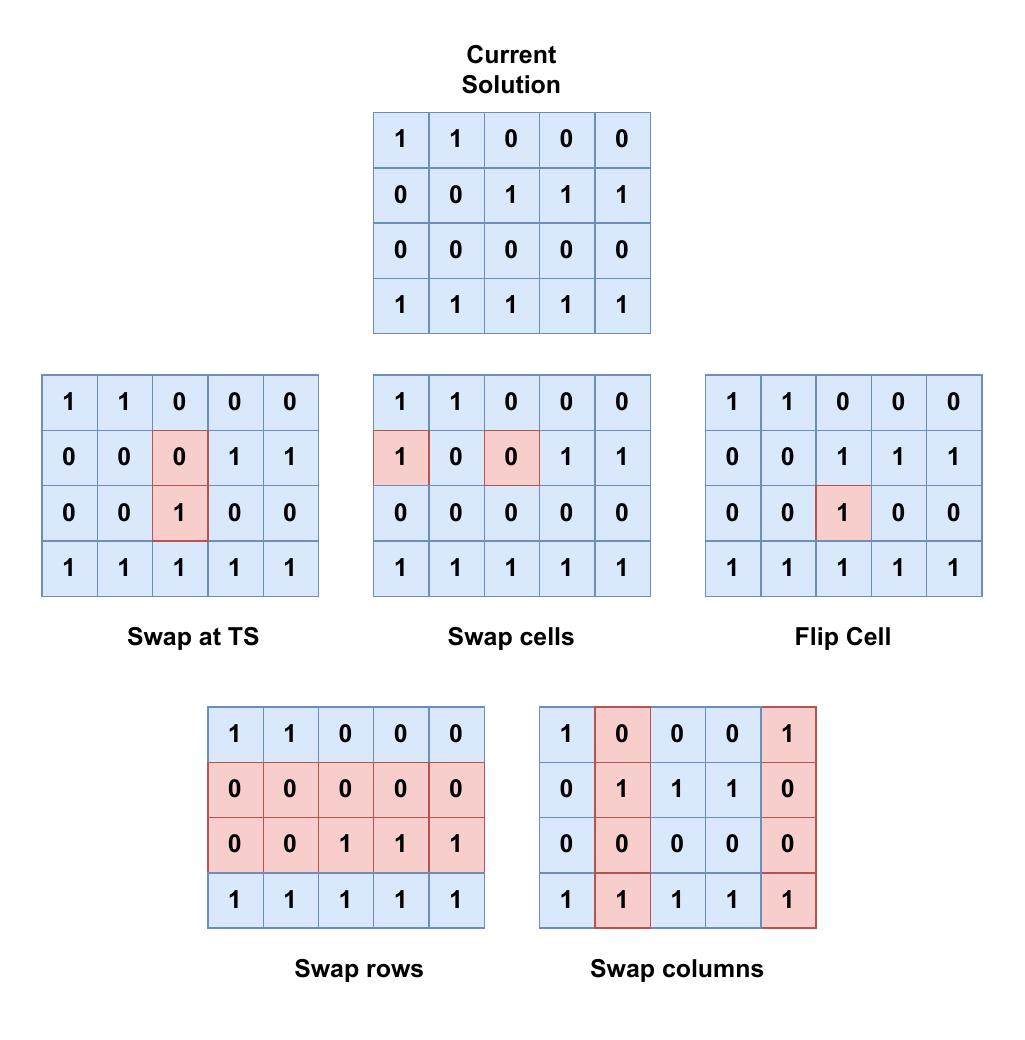}
    \caption{Overview with examples of the operations with which neighbor solutions are created in the simulated annealing approach.}
    \label{fig:sa_operations}
\end{figure}

How neighbors are constructed is problem-specific. In this work, neighboring solutions (i.e., the qubit schedule matrix) are created using one of the methods illustrated in Fig. \ref{fig:sa_operations}; which method is used is determined randomly in each iteration. The initial solution is constructed by filling each QPU successively with consecutive qubits.

\begin{algorithm}[tb]
    \caption{High level overview of simulated annealing}
    \label{alg:simulated_annealing}
    \begin{algorithmic}
    \REQUIRE $max\_iterations$
    \REQUIRE $initial\_temp$ 
    \REQUIRE $cooling\_rate$ 
    \REQUIRE $initial\_solution$
    \STATE $current\_solution \leftarrow initial\_solution$
    \STATE $best\_solution \leftarrow current\_solution$ 
    \STATE $cc \leftarrow calculate\_cost(current\_solution) $ \COMMENT{Current cost}
    \STATE $bc \leftarrow calculate\_cost(best\_solution)$ \COMMENT{Best Cost found}
    \FOR{$i \leftarrow 0$ to $max\_iterations$}
         \STATE $temp \leftarrow temp\ *\ cooling\_rate$
         \STATE $n \leftarrow get\_neighbor(current\_solution)$ 
         \STATE $nc \leftarrow calculate\_cost(neighbor) $
         \IF{nc $<$ cc\ or accept\_prob(cc, nc, temp) $>$ random(0, 1)  }
            \STATE $current\_solution \leftarrow neighbor$
            \STATE $cc\leftarrow nc$
        \ENDIF
        \IF{cc $<$ bc}
            \STATE $best\_solution \leftarrow current\_solution$
            \STATE $bc \leftarrow cc$
        \ENDIF
    \ENDFOR
    \RETURN $best\_solution$
    \end{algorithmic}
\end{algorithm}

\subsubsection{\textbf{Evolutionary Algorithm}}
Evolutionary Algorithms (EAs) \cite{eiben2002evolutionary,holland1992adaptation} are meta-heuristic optimization algorithms that can be applied to numerous problems in a variety of domains. Various algorithms fall under the umbrella of EAs -- all utilizing some aspects inspired by evolution -- however, we will be focusing on the main points in this section and refrain from delving into the details of each approach. EAs are inspired by nature, more specifically biological Darwinian evolution, in that they mimic some aspects of the natural world, albeit in a simplistic form, to solve optimization problems. This is achieved by successively combining and refining existing solutions in order to incrementally improve individuals and move closer towards the global optimum. That is, an EA consists of a \textit{population} of \textit{individuals}, i.e., solutions, where each individual has its own corresponding \textit{fitness}, i.e., a scalar value indicating how good a particular solution is. The algorithm runs for a number of \textit{generations}, and in each generation new individuals are created by combining components of existing ones in a process known as \textit{crossover}. Individuals can be subjected to random \textit{mutations}, a process in which a small aspect of a solution is randomly adjusted. Note that the fitness function, solution representation as well as crossover and mutation operations are all problem specific, i.e., must be implemented and tailored to each problem individually, however, some aspects or techniques can be re-used from other problems.
We now discuss the first of two EAs in this work, which is based on \cite{sunkel2024applying}, and we refer to this one as the ``EA Scheduler'', as it optimizes the qubit schedule without changing the underlying circuit. We will present the main components of this approach next.
An individual (i.e., solution) is initialized in one of two ways, which one is used is determined randomly. In the first solution initialization method, each qubit is assigned to a single QPU for every time step, each QPU being successively filled up, and once it is full, the next QPU is used. After the solution is created, it is randomly shuffled (i.e., the rows of the solution matrix are shuffled) to increase the diversity in the population. In the second solution initialization method, every element of the solution is determined randomly.

The EA uses two different versions of the single-point crossover method, and we refer to them as single-point row-wise and single-point column-wise crossover. In both methods, a random crossover point is selected that determines which ``genes'' come from each parent. With a single cut-off point this is straightforward; all genes from one side of the point come from on parent while the rest from the other parent. However, as the solution in our EA is a 2 dimensional list, we employ two different variants of this method. In the first approach, all ``genes'' left to the randomly selected crossover point are taken from parent 1 while the right side is taken from parent 2. In the second crossover method on the other hand, all ``genes'' above the crossover point are taken from parent 1 while the rest are taken from parent 2.

Various mutation methods are applied throughout the evolutionary process. The \textit{single flip mutation} selects a random element from the solution and changes its value, i.e., assigns a qubit to a different QPU at a randomly chosen time point. A variant of this, the \textit{multiple flip mutation}, applies the previous mutation method several times in succession, the number of repetitions is randomly determined. In the \textit{swap rows mutation}, two rows are randomly selected and their values are swapped, whereas in the \textit{swap columns mutation} the same is applied to entire columns. In the \textit{swap nodes mutation}, two nodes at a particular time step are swapped. In the \textit{shuffle subset rows mutation}, random start and end points are selected, and the rows between these points are randomly shuffled; analog behaviour for columns is provided through the \textit{shuffle subset columns mutation}. What mutations are applied is also randomly determined, i.e., not all mutations are necessarily applied in each generation.

\subsubsection{\textbf{Baselines}}
We compare the approaches to the following three different baselines, namely graph partitioning, a sequential and random sequential assignment.

\paragraph*{\textbf{Graph Partitioning}}
Assigning qubits to QPUs in a quantum network can be formulated as a graph partitioning problem, and various approaches based on this idea have been evaluated by the research community, cf. Sec. \ref{sec:related_work}. The graph partitioning (GP) problem can be defined as follows: Given a graph $G = (N, E)$ where $N$ is a set of nodes and $E$ a set of edges, partition the graph into two distinct subsets such that each contains an equal number of nodes while minimizing the number of edges between the partitions. The problem can be extended such that the graph G is partitioned into $k$ subsets, where $k \geq 2$. 
In the context of minimizing teleportation costs in DQC, a circuit can be transformed into a graph where the nodes are qubits and edges represent two qubit gates (e.g., CX). Edges are weighted according to the number of CXs between the respective qubits. The GP algorithm can then be applied on this graph in order to find an assignment of qubits to QPUs while minimizing two gate operations between qubits in different partitions. 
In this work, we use PyMetis \cite{pymetis}, which is based on METIS \cite{karypis1998fast}, for the GP implementation. 
Furthermore, the resulting GP partitioning is then converted to the schedule mentioned above; the cost is calculated based on this converted assignment. Because the GP algorithm cannot consider changing partitions over time, each qubit is assigned the respective QPU according to the static partitioning for all time steps of the schedule.

\paragraph*{\textbf{Sequential assignment}}
In this assignment strategy, each QPU is assigned successive qubits until its capacity is reached and the assignment does not change over time, i.e., the schedule remains static. For example, in a network with 4 nodes and a circuit containing 8 qubits, the first QPU is assigned qubits 0 and 1, the next QPU 2 and 3, and so on.

\paragraph*{\textbf{Random sequential assignment}}
This strategy is similar to the previous, the only difference being that the rows are shuffled. That is, each QPU is still filled to its maximum capacity, and the assignment remains static. However, what qubit is assigned to each QPU is random; though each qubit can only be assigned to one QPU.

\subsection{Optimizing the Circuit}
We next introduce our approach in which the circuit itself is optimized instead of the qubit assignment schedule, and we refer to this as the quantum circuit optimization (QCO) approach. For this, we propose an evolutionary algorithm that can adjust a given circuit guided by a fitness function. We begin with a high-level description of the objective before giving a formal definition of the fitness function; we then introduce the algorithm itself.
In this work, we focus on a state preparation task and, therefore, the first objective is to maximize the fidelity $F$ of the state prepared by the optimized circuit to the one produced by the original. More formally \cite{ge2024quantum}:

\begin{equation}
    \max F(\ket{\psi}, U) = |\bra{\psi}U\ket{0}|^2
\end{equation}

\noindent where $\ket{\psi}$ is the target state and $U$ a unitary, i.e., circuit. 

The second part of the fitness function should reduce the communication cost in the same vain as the algorithms in the previous section. That is, based on the optimized circuit, a qubit schedule is determined, which is then used to calculate the communication cost for that particular circuit and qubit assignment, i.e., $u(\hat{U}) = f(p(\hat{U})$ where $f$ is the cost function (cf. Eq. \ref{eq:communication_cost}) and $p$ a qubit assignment algorithm that determines the schedule defined above. The schedule is determined by applying the PyMetis algorithm and converting the partitioning to the formulation, however, in principle the other algorithms discussed above could also be used. The resulting communication cost is then normalized.
The overall fitness function is defined as:

\begin{equation}
    \max_{\hat{U}} g(U, \hat{U}) = F(U\ket{0}, \hat{U}) - \frac{u(\hat{U})}{u(U)}
\end{equation}

\noindent where $U$ is the original circuit, $\hat{U}$ the optimized circuit, $F(\ket{\psi}, U) = |\bra{\psi}U\ket{0}|^2$ calculates the fidelity for the state preparation task, and $u$ the communication cost of a given circuit, or more specifically, of the corresponding schedule. Solutions below a fidelity threshold are penalized, so the fitness employed in the EA is as follows:

\begin{equation}
    g(U, \hat{U}) = \begin{cases}
         \lambda & \text{if } F(U\ket{0}, \hat{U}) < 1 - \epsilon \\
         F(U\ket{0}, \hat{U}) - \frac{u(\hat{U})}{u(U)} & \text{else}
    \end{cases}
\end{equation}

\noindent where $\epsilon$ was set to $0.003$ and $\lambda$ to $-100$; this is to ensure that the solutions retain a high fidelity to the target state.

The EA to optimize the circuit is structured as follows. A circuit is represented by a one-dimensional list of gates. Each gate is an object containing all relevant information such as gate name (e.g., ``X'', ``RZ'', etc.), qubits it acts on, or parameters. The evolutionary operations (i.e., mutation and crossover) are performed on the solution representation; the solution is then converted to a quantum circuit from which the fidelity can then be calculated. Offspring, i.e., new individuals are created via single-point or uniform crossover; which method is used is determined randomly with equal probability. Whether crossover is utilized to create new solutions is determined by the crossover rate hyperparamter. Alternatively, individuals can be initialized with the original circuit in the same way when the initial population is created. Individuals are subject to further mutation according to a mutation rate that determines if a mutation method should be applied. Each individual is mutated at most once, and the mutation method used is determined randomly. The EA uses the following mutation methods: ``add gate'', ``remove gate'', ``swap gates'', ``shuffle subset'', and ``mutate gate''. The ``shuffle subset'' mutation randomly selects two points in the solution list and shuffles the gates, and the ``mutate gate'' changes an existing gate. The EA follows an elitist approach, that is, the worst $n$ individuals are replaced by $m$ children, i.e., the best individuals are guaranteed to be transferred to the new generation.

\begin{figure*}[tb]
    \centering
    
    \scalebox{0.8}{
    \begin{subfigure}[b]{1\textwidth}
        \includegraphics[width=\textwidth]{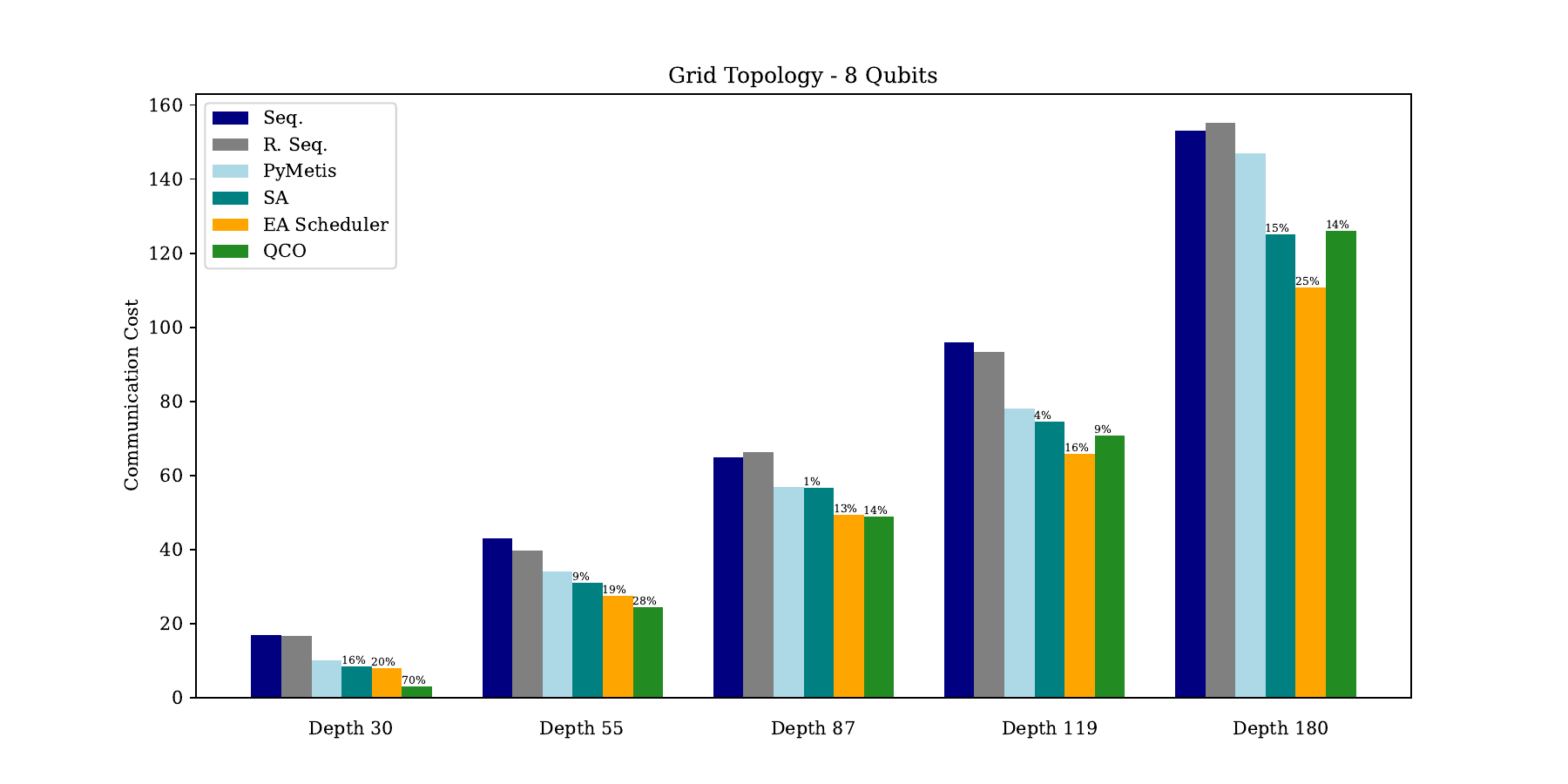}
        \caption{Grid topology with 4 nodes.}
        \label{fig:8_qubits_4_nodes_grid_bar_plot}
    \end{subfigure}}
    
    \scalebox{0.8}{
    \begin{subfigure}[b]{1\textwidth}
        \includegraphics[width=\textwidth]{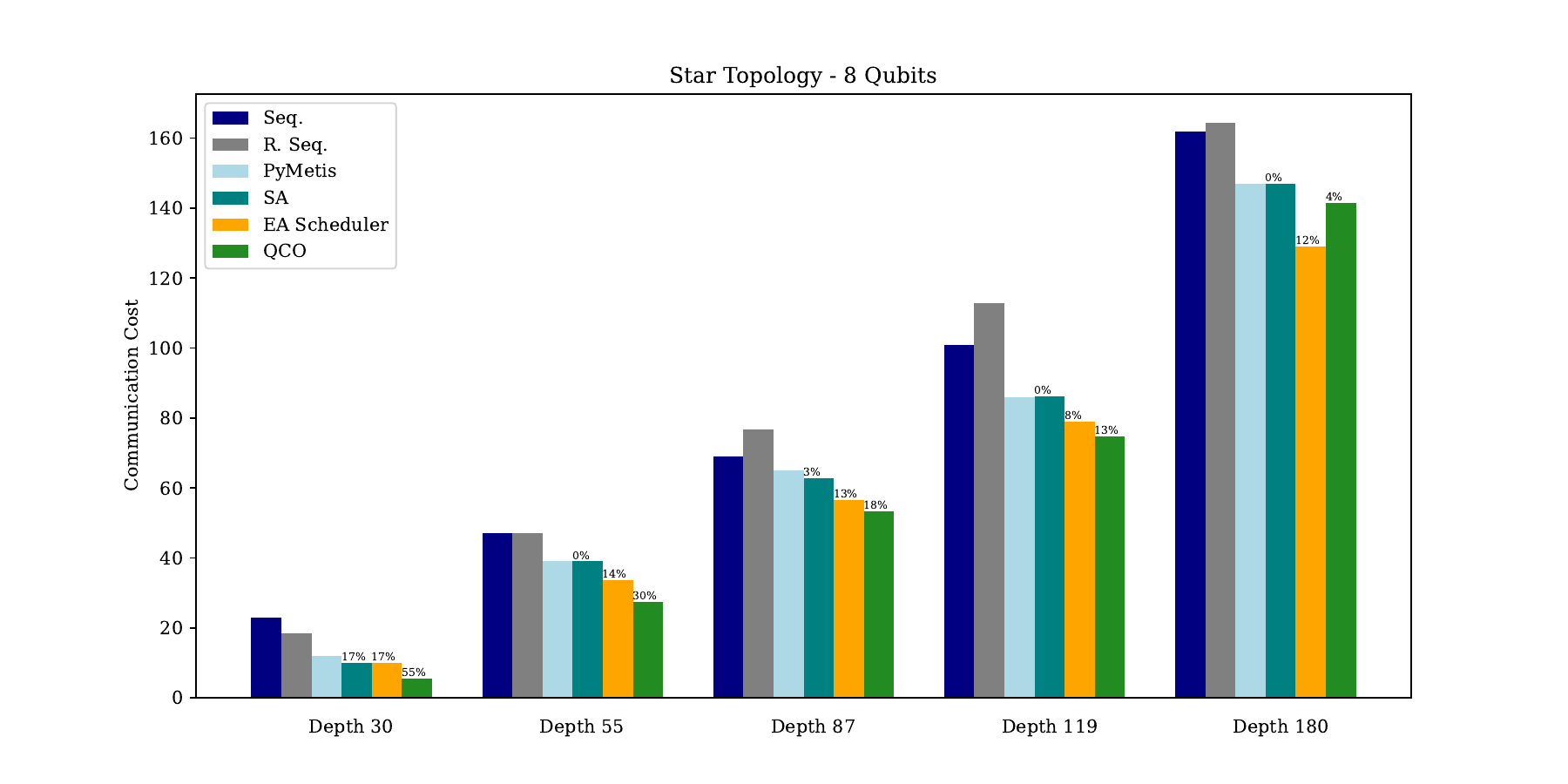}
        \caption{Star topology with 4 nodes.}
        \label{fig:8_qubits_4_nodes_star_bar_plot}
    \end{subfigure}}
    \caption{Bar plots showing the communication cost for different topologies and circuits containing $8$ qubits. Numbers above bars indicate the improvement over the GP baseline.}
\end{figure*}

\begin{figure*}[tb]
    \centering
    \scalebox{0.8}{\includegraphics[width=\textwidth]{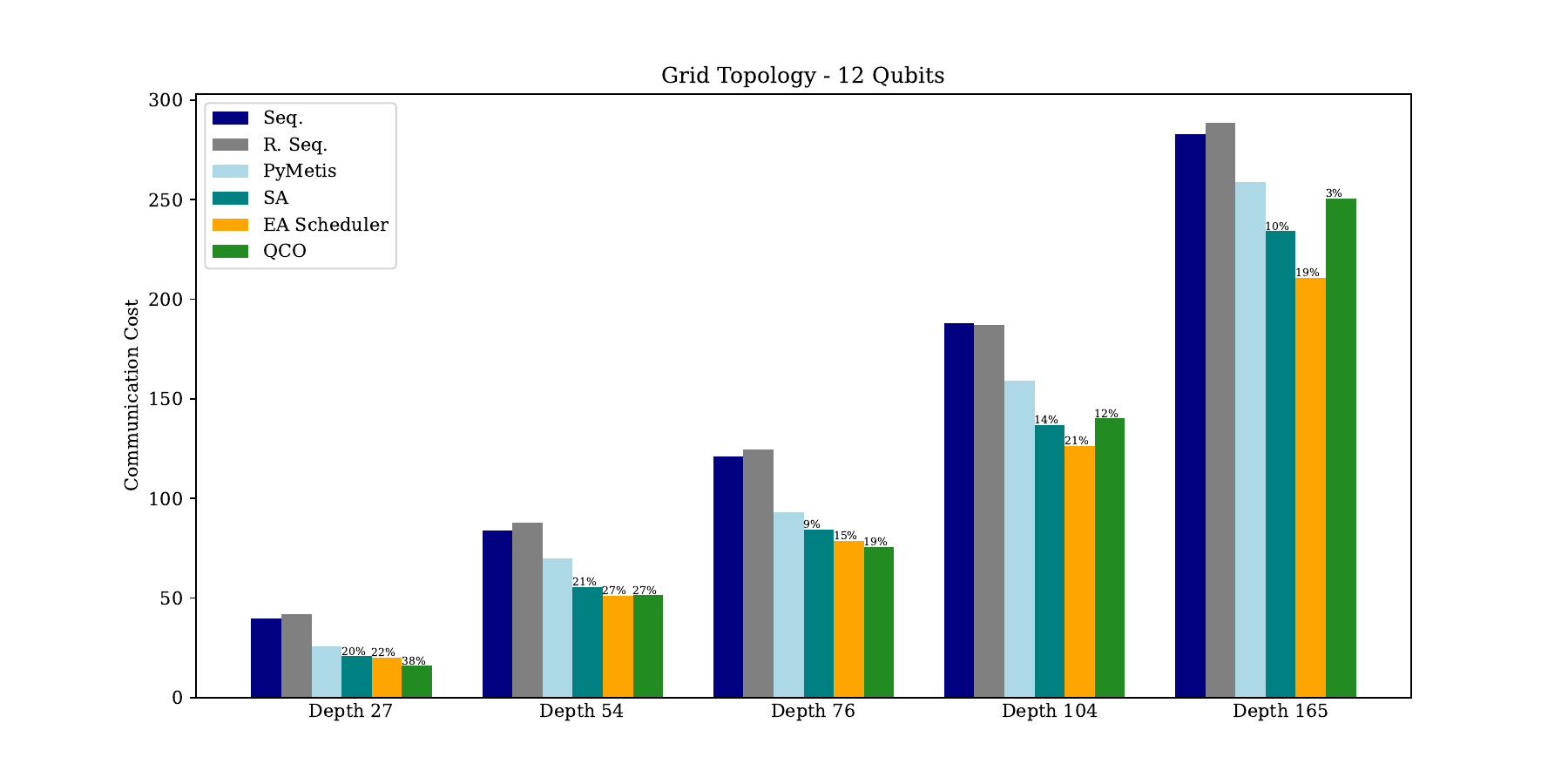}}
    \caption{Grid topology with $6$ nodes.}
    \label{fig:12_qubits_6_nodes_grid_bar_plot}
\end{figure*}

\begin{figure*}
    \centering
    \scalebox{0.8}{\includegraphics[width=\textwidth]{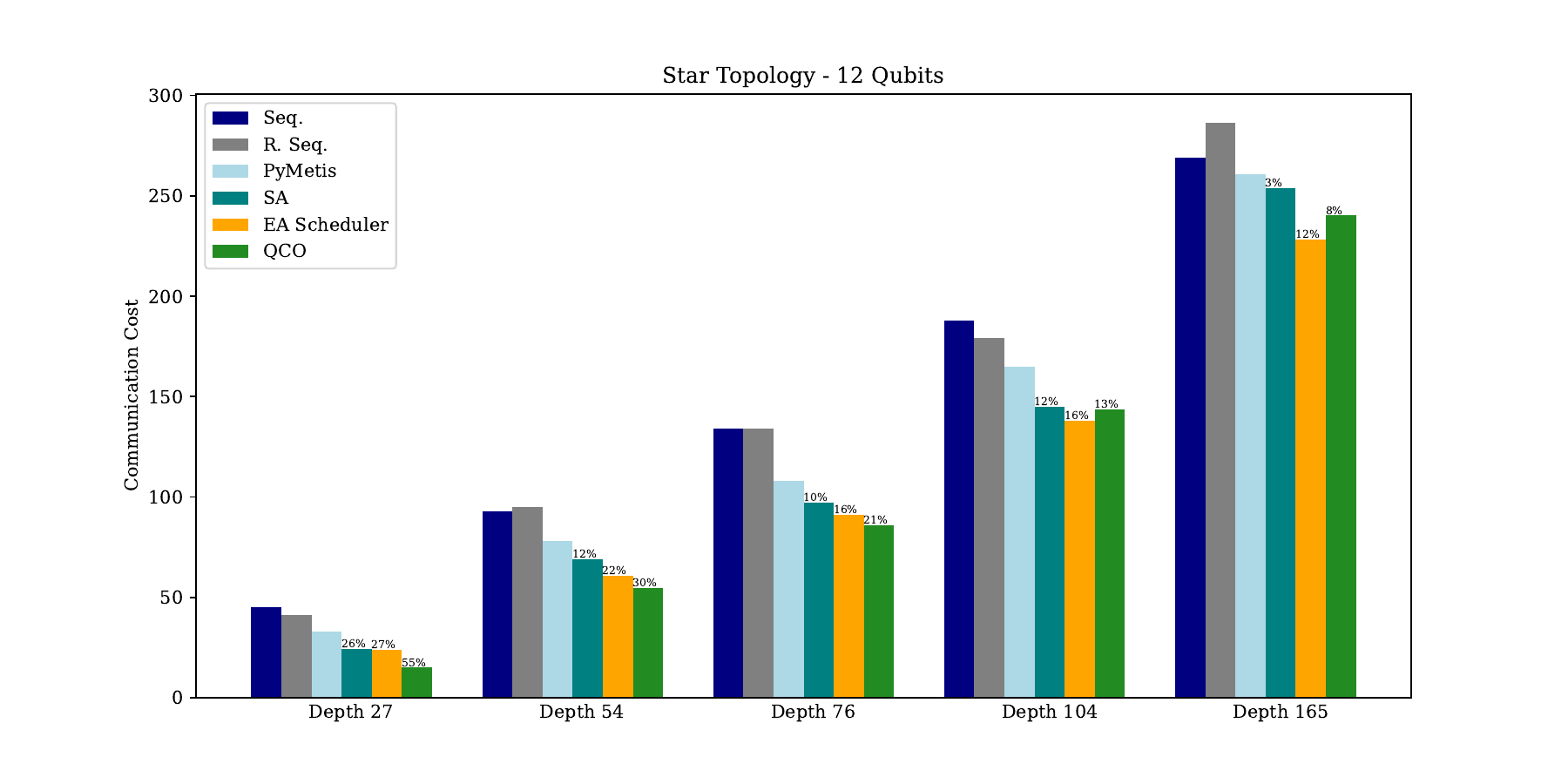}}
    \caption{Star topology with $6$ nodes.}
    \label{fig:12_qubits_6_nodes_star_bar_plot}
\end{figure*}

\begin{figure*}[tb]
    \centering
    \scalebox{1}{
    \begin{subfigure}[b]{0.45\textwidth}
        \centering
        \includegraphics[width=\textwidth]{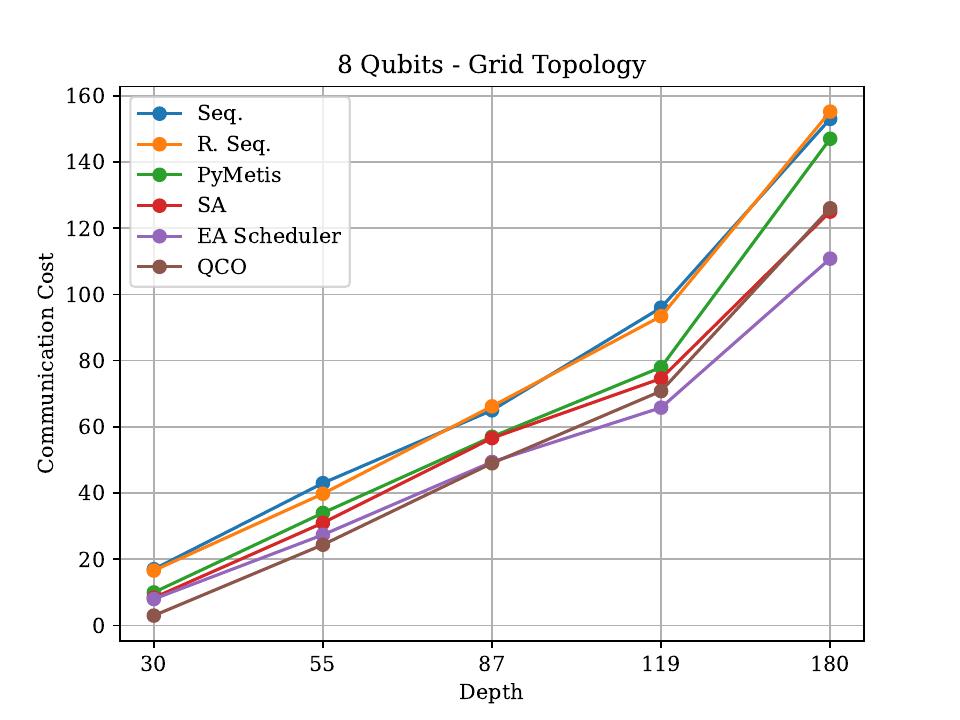}
        \caption{Grid topology}
        \label{fig:8_qubits_4_nodes_grid_depth_plot}
    \end{subfigure}}
    \hfill
    \scalebox{1}{
    \begin{subfigure}[b]{0.45\textwidth}
        \centering
        \includegraphics[width=\textwidth]{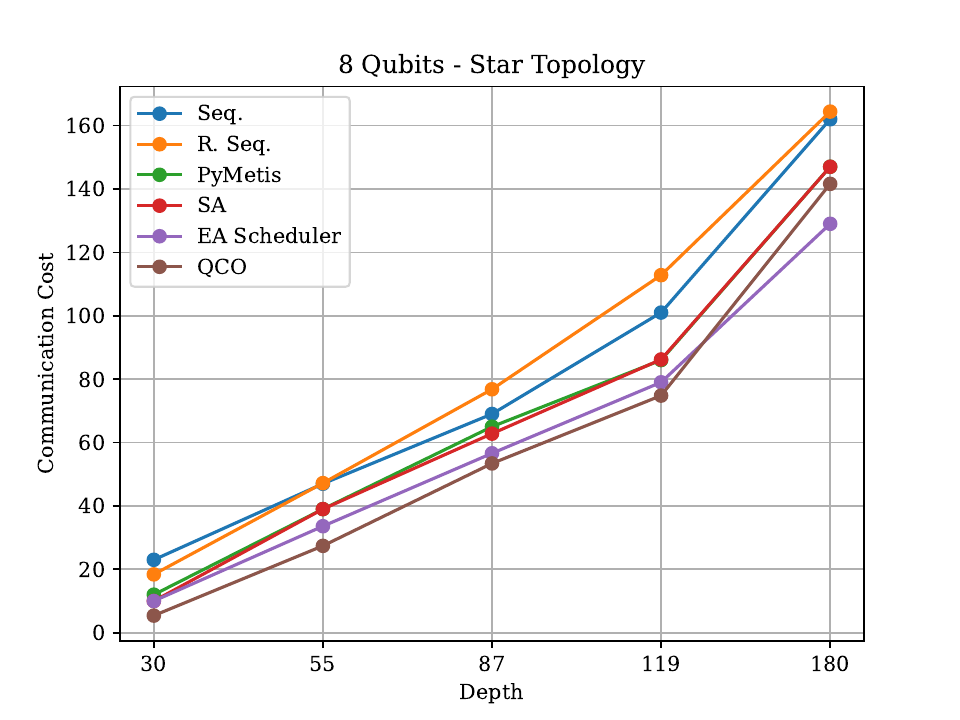}
        \caption{Star topology}
        \label{fig:8_qubits_4_nodes_star_depth_plot}
    \end{subfigure}}
    \scalebox{1}{
    \begin{subfigure}[b]{0.45\textwidth}
        \centering
        \includegraphics[width=\textwidth]{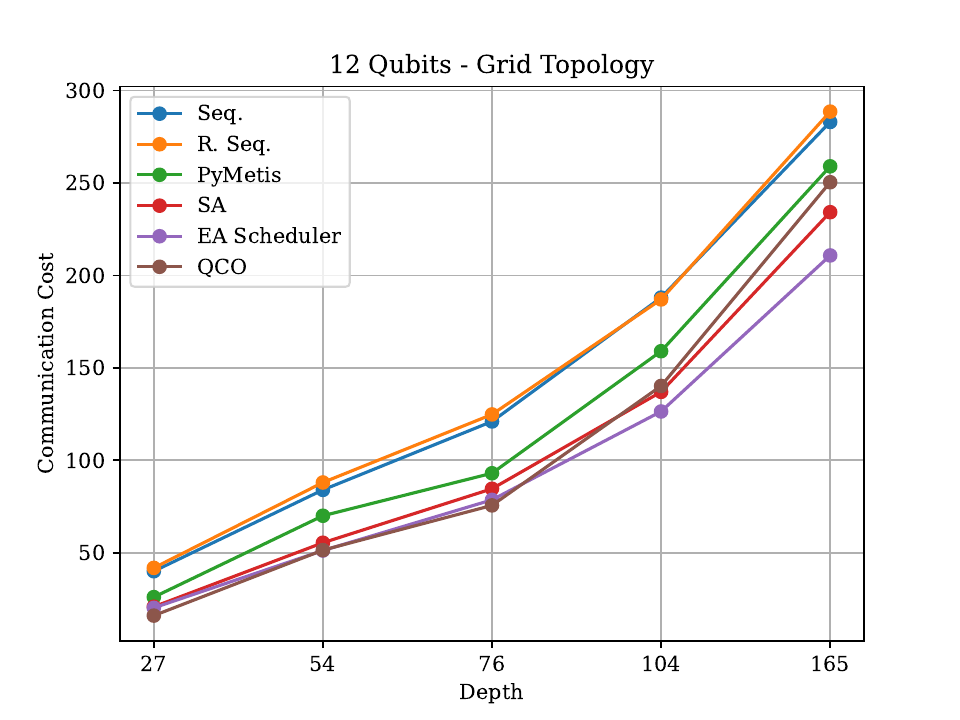}
        \caption{Grid topology}
        \label{fig:12_qubits_6_nodes_grid_depth_plot}
    \end{subfigure}}
    \hfill
    \scalebox{1}{
    \begin{subfigure}[b]{0.45\textwidth}
        \centering
        \includegraphics[width=\textwidth]{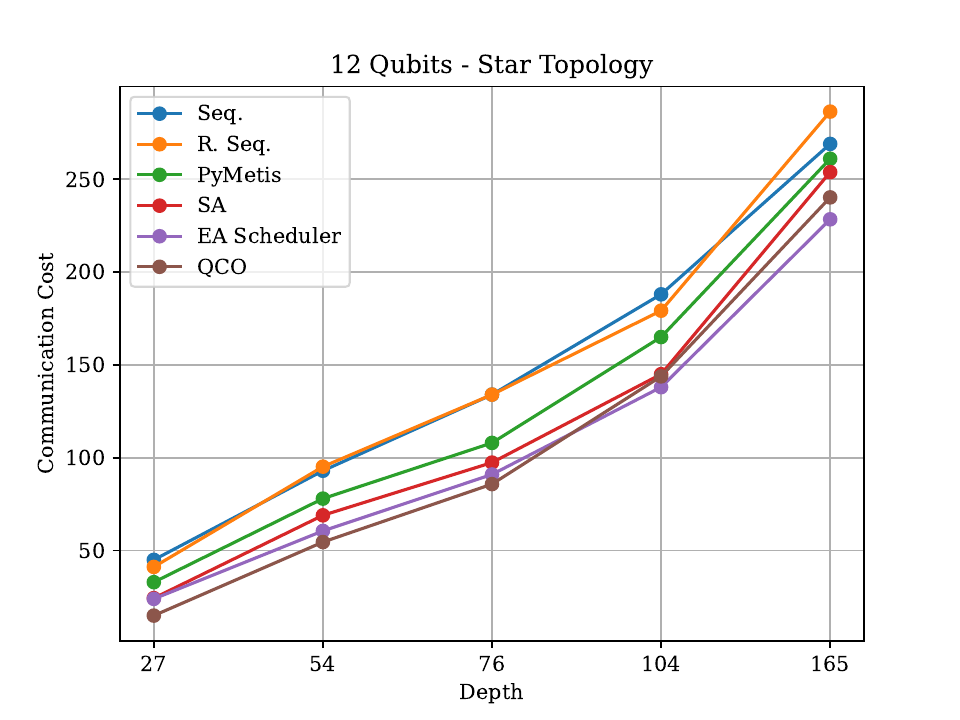}
        \caption{Star topology}
        \label{fig:12_qubits_6_nodes_star_depth_plot}
    \end{subfigure}}
    \caption{Networks with $4$ nodes and circuits with 8 (top) and $12$ (bottom) qubits.}
    \label{fig:enter-label}
\end{figure*}

\begin{figure*}[tb]
    \centering
     \scalebox{1}{
    \begin{subfigure}[b]{0.45\textwidth}
        \centering
        \includegraphics[width=\textwidth]{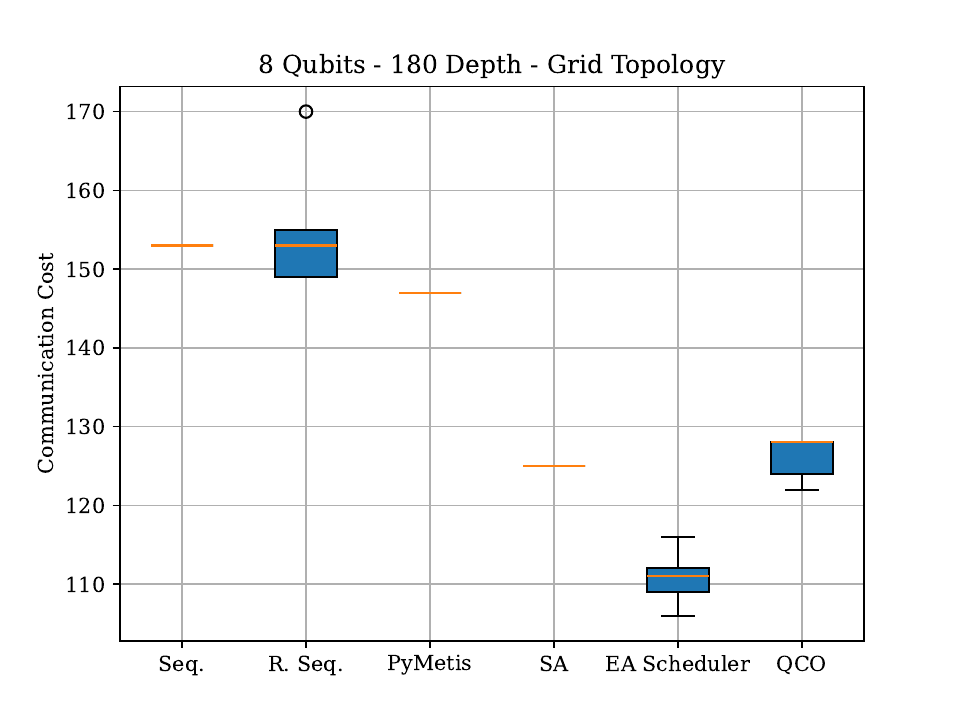}
        \caption{Grid topology with $4$ nodes.}
        \label{fig:8_qubits_4_nodes_grid_box_plot}
    \end{subfigure}}
    \hfill
    \scalebox{1}{
    \begin{subfigure}[b]{0.45\textwidth}
        \centering
        \includegraphics[width=\textwidth]{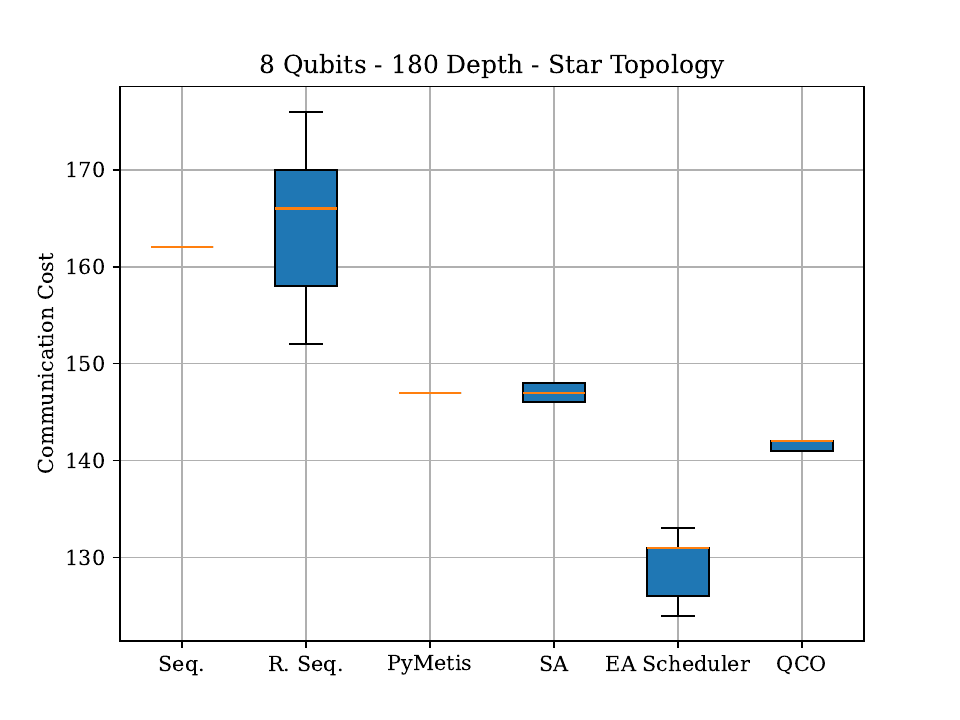}
        \caption{Star topology with $4$ nodes.}
         \label{fig:8_qubits_4_nodes_star_box_plot}
    \end{subfigure}}
    \scalebox{1}{
    \begin{subfigure}[b]{0.45\textwidth}
        \centering
        \includegraphics[width=\textwidth]{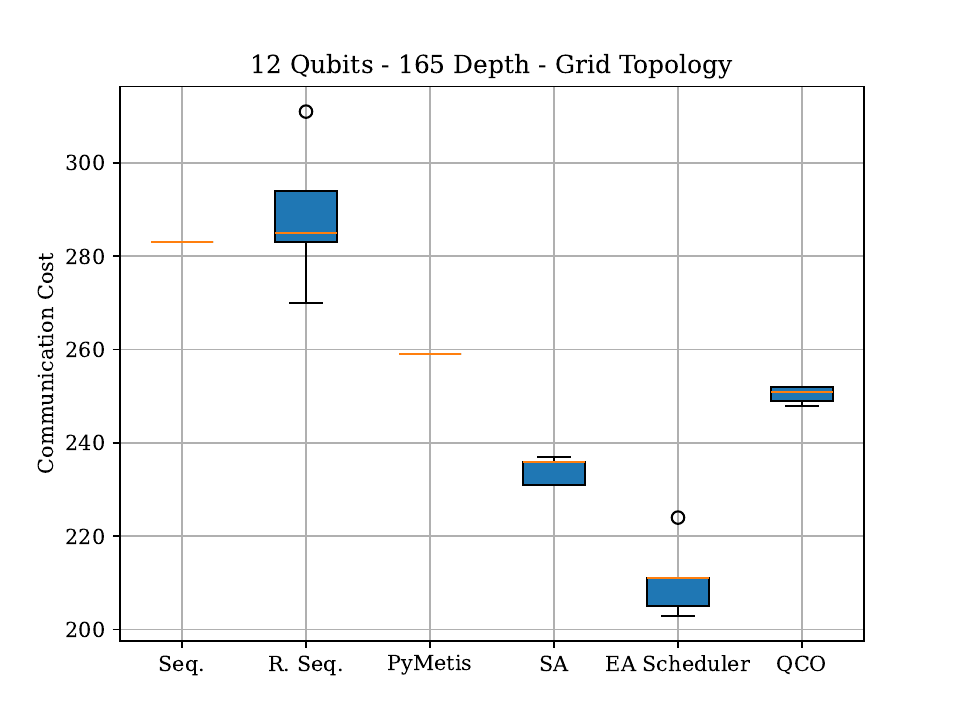}
        \caption{Grid topology with $6$ nodes.}
        \label{fig:12_qubits_6_nodes_grid_box_plot}
    \end{subfigure}}
    \hfill
    \scalebox{1}{
    \begin{subfigure}[b]{0.45\textwidth}
        \centering
        \includegraphics[width=\textwidth]{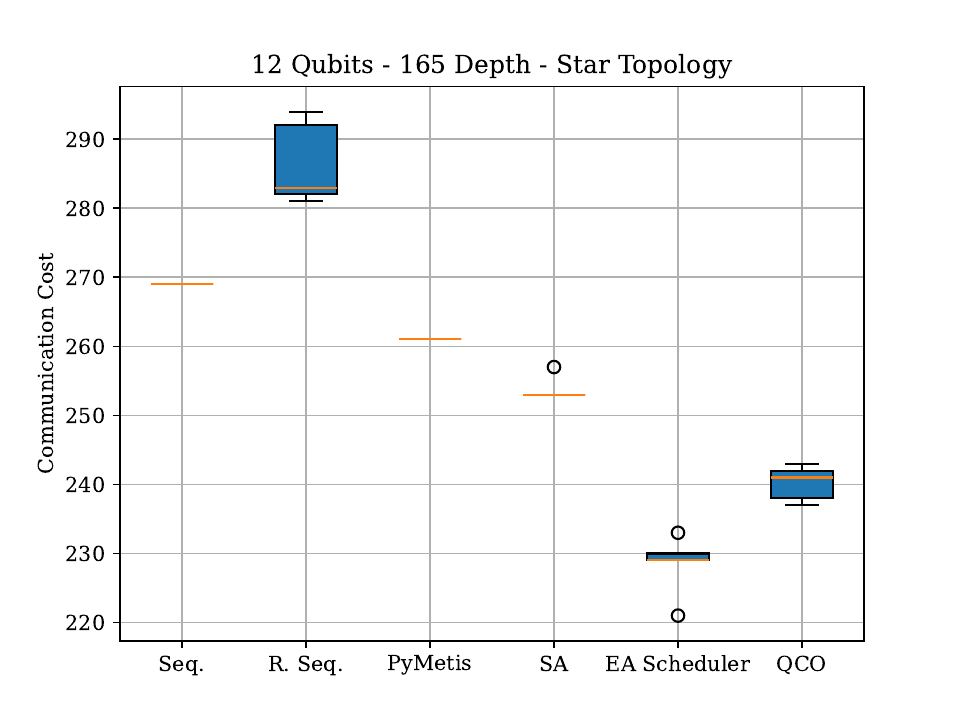}
        \caption{Star topology with $4$ nodes.}
        \label{fig:12_qubits_6_nodes_star_box_plot}
    \end{subfigure}}
    \caption{Networks with $4$ nodes and circuits with $8$ qubits.}
    \label{fig:8_qubits_4_nodes_box_plot}
\end{figure*}

\section{Experimental Setup}\label{sec:experimental_setup}
In this section, we will first describe our experimental setup, i.e., we list circuits, network topologies, and the configuration as well as parameters for the algorithms. 
We use Qiskit \cite{Qiskit} to create various random circuits with $8$ and $12$ qubits with increasing depth and thus difficulty. Each circuit was further transpiled, and CX gates are the only multi-qubit gates present in the circuit, the depth shown in the figures corresponds to the circuit after transpilation using the following gate set $[X, SX, CX, RZ]$.
All experiments were run with $5$ different seeds. We describe the parameters of the EAs and SA next. The \textit{population size} is the total number of individuals in the population. The \textit{crossover rate} determines the probability that a child is created through crossover, children can alternatively be initializing a new individual unrelated to a parent, which should increase the diversity in the population. The \textit{mutation rate} is the probability that a random mutation is performed on a child after it has been created. \textit{Offspring rate} is the number of children created in each generation while \textit{replace rate} determines the number of individuals in the population that are replaced by the children, both are the percentage of the total population size. Note that for the QCO approach, new circuits are created with up to $100$ gates. SA utilizes the following parameters: the start temperature and the cooling rate, and number of iterations which was set to $100000$. However, the temperature can not drop below $0.00001$. The above parameters used were determined experimentally using Optuna \cite{optuna_2019} with a different configuration for each topology and network size. 
We used two network topologies in the experiments ($4$ and $6$ nodes), that is, a grid and a star network where each node represents a QPU with a capacity of $2$ qubits.

\section{Results}\label{sec:numerical_results}
The results of all experiments are discussed in this section. A comparison of the communication cost for the experiments with circuits containing 8 qubits is shown in Fig. \ref{fig:8_qubits_4_nodes_grid_bar_plot} and Fig. \ref{fig:8_qubits_4_nodes_star_bar_plot} for the grid and star networks with 4 nodes, respectively. Note that the numbers in the plot indicate the improvement of the EAs and SA over the GP baseline. In the grid topology, the QCO approach achieves the best cost reduction for the first $3$ circuits; however, the rate decreases with increasing depth. For example, in the circuit with a depth of $30$, a cost reduction of $70$\% compared to the GP baseline is achieved; however, this decreases to 14\% in the largest circuit with a depth of $180$. SA and the EA Scheduler on the other hand achieve their best results on the largest circuit, with a decrease by $15$\% and $25$\% respectively. However, in each instance, either EA achieves the best results. 
With the star topology a similar picture emerges where QCO is able to deliver the best results on $4$ of the $5$ circuits; on the largest circuit, however, it is only able to achieve a $4$\% improvement over GP. SA is only able to achieve an improvement on $2$ circuits with 17\% on the smallest and $3$\% on the middle circuit with the rest achieving the same results as the GP baseline. The EA Scheduler achieves the best results on the largest circuit while achieving improvements on all other circuits ranging from $8$\% to $17$\%.
The scaling behavior is depicted in Fig. \ref{fig:8_qubits_4_nodes_grid_depth_plot} and Fig.\ref{fig:8_qubits_4_nodes_star_depth_plot}, which shows that with larger circuits, the difference between the approaches is more evident. Fig. \ref{fig:8_qubits_4_nodes_grid_box_plot} and \ref{fig:8_qubits_4_nodes_star_box_plot} show the box plots with the results of the circuits with a depth of 180; showing that with larger circuits the variance is evident.

The results for the experiments with $12$ qubit circuits are shown in Fig. \ref{fig:12_qubits_6_nodes_grid_bar_plot} and Fig. \ref{fig:12_qubits_6_nodes_star_bar_plot} for the grid and star topologies respectively. How the algorithms scale can be seen in Fig. \ref{fig:12_qubits_6_nodes_grid_depth_plot} and Fig. \ref{fig:12_qubits_6_nodes_star_depth_plot} and the variance of different runs in figures \ref{fig:12_qubits_6_nodes_grid_box_plot} and \ref{fig:12_qubits_6_nodes_star_box_plot}. The results with $12$ qubits follow similar patterns as with the smaller circuits. In the grid topology, the QCO approach achieves the best results for two circuits and is equal to the EA scheduler in one circuit. For the two largest circuits the EA scheduler achieves the best cost reduction. In the star topology, QCO achieves the best results in three out of five circuits while the EA scheduler again achieves the best results on the two largest circuits.

\section{Discussion}\label{sec:discussion}
The proposed approaches, in particular the EAs, provide significant improvements over the baselines. The main advantage of the problem formulation is the fact that it incorporates both the changing connectivity between qubits over time as well as the underlying network topology whereas the GP algorithm, for example, is not able to process the temporal aspect, as it is lost if the entire circuit is converted into a graph. The EA Scheduler in particular is able to optimize the schedule in this regard; deciding when to apply teleportation and when a remote CX operation. The downside of this approach is that it scales with both the number of qubits and the circuit's depth, making it difficult and time-consuming to optimize larger circuits. The GP approach is more efficient in this regard and maybe sufficient if fast results are required that are better than a naive sequential assignment. Our second EA, the QCO approach that optimizes the circuit rather than the schedule directly, is also able to achieve better results than the baselines. However, as with the EA Scheduler, it is a more time-consuming approach that scales with the circuit's size. The QCO as presented is suitable for state preparation tasks and can achieve significant reduction in communication costs. The balance between required fidelity and communication cost requires further study though, as this might be dependent on the application. In this work, high fidelity was required and constrained the optimization. If, however, other applications allow a more lenient approach, further reduction in communication may be possible, this, however, is left for future work. With increased circuit size, the improvement rate for the QCO approach decreased; this should also be investigated in further work. It is possible that through running the algorithm for more generations or by increasing the size of the initialized circuits could lead to improved results. This, however, would increase the computational time. The method could also be seen as a preprocessing step in a DQC compilation pipeline before other scheduling or qubit assignments are applied..

\section{Conclusion}\label{sec:conclusion}
DQC over quantum networks promises to scale quantum computing to new dimensions; however, it does come with its own challenges due to communication overhead necessary to enable large-distance quantum communication. Novel optimization problems emerge as well as the need for a new form of compilers. We approached the problem of qubit assignment to QPUs in a network from two directions by optimizing the schedule through time-aware algorithms, and by applying an evolutionary-based quantum circuit optimization algorithm to adjust the circuit itself rather than the schedule to minimize non-local operations. Both EAs are able to significantly reduce the communication cost compared to the baselines, ranging from $13$\% to $70$\% in cost reduction compared to GP. Our SA and EA schedule optimization techniques take advantage of considering the temporal-dimension of the circuits as well as the network topology which enables to achieve improvements over other methods (e.g. GP) that do not consider this information. The proposed methods could be integrated into compilation stacks for DQC. The QCO approach in particular could be used to preprocess circuits such that they optimally suit a specific network topology and should be investigated further.

\bibliographystyle{IEEEtran}
\bibliography{IEEEabrv,bibliography}

\end{document}